\documentclass[sigconf]{acmart}

\usepackage{xspace}
\usepackage{tabularx}
\usepackage{alltt}
\usepackage{geometry}
\usepackage{array}
\usepackage{enumitem}
\usepackage{caption}
\usepackage{subcaption}
\usepackage{ragged2e}

\AtBeginDocument{%
  \providecommand\BibTeX{{%
    \normalfont B\kern-0.5em{\scshape i\kern-0.25em b}\kern-0.8em\TeX}}}

\copyrightyear{2022} 
\acmYear{2022} 
\setcopyright{rightsretained} 
\acmConference[ASSETS '22]{The 24th International ACM SIGACCESS Conference on Computers and Accessibility}{October 23--26, 2022}{Athens, Greece}
\acmBooktitle{The 24th International ACM SIGACCESS Conference on Computers and Accessibility (ASSETS '22), October 23--26, 2022, Athens, Greece}\acmDOI{10.1145/3517428.3544796}
\acmISBN{978-1-4503-9258-7/22/10}

\newcommand\blue[1]{{\color{blue}#1}}

\newcommand\percleveltwo{{\color{black}{50\%}}\xspace}
\newcommand\perclevelthree{{\color{black}{31\%}}\xspace}
\newcommand\totalalttext{{\color{black}{547}}\xspace}
\newcommand\totalalttextsents{{\color{black}{2127}}\xspace}

\newcommand\numtotalpdfs{{\color{black}{25218}}\xspace}
\newcommand\numconvertedpdfs{{\color{black}{19500}}\xspace}
\newcommand\percconvertedpdfs{{\color{black}{77.3\%}}\xspace}
\newcommand\numvalidalttext{{\color{black}{3386}}\xspace}
\newcommand\numpaperswithvalidalttext{{\color{black}{897}}\xspace}
\newcommand\percpaperswithvalid{{\color{black}{4.6\%}}\xspace}
\newcommand\numannotated{{\color{black}{1085}}\xspace}

\newcommand\githublink{\href{https://github.com/allenai/hci-alt-texts}{https://github.com/allenai/hci-alt-texts}\xspace}

\newcolumntype{L}[1]{>{\raggedright\let\newline\\\arraybackslash\hspace{0pt}}p{#1}}
\newcolumntype{M}[1]{>{\raggedright\let\newline\\\arraybackslash\hspace{0pt}}m{#1}}

\definecolor{darkgreen}{rgb}{0.0, 0.4, 0.13}

\begin{document}

\title{A Dataset of Alt Texts from HCI Publications}
\subtitle{Analyses and Uses Towards Producing More Descriptive Alt Texts of Data Visualizations in Scientific Papers}

\author{Sanjana Chintalapati}
\authornote{Work conducted during internship at Allen Institute for AI}
\email{sanjanac@cs.washington.edu}
\orcid{}
\affiliation{%
  \institution{University of Washington}
  \city{Seattle}
  \state{WA}
  \postcode{98103}
  \country{USA}
}

\author{Jonathan Bragg}
\email{jbragg@allenai.org}
\orcid{}
\affiliation{%
  \institution{Allen Institute for AI}
  \city{Seattle}
  \state{WA}
  \postcode{98103}
  \country{USA}
}

\author{Lucy Lu Wang}
\email{lucylw@uw.edu}
\orcid{0000-0001-8752-6635}
\affiliation{%
  \institution{University of Washington; \\Allen Institute for AI}
  \city{Seattle}
  \state{WA}
  \postcode{98103}
  \country{USA}
}

\renewcommand{\shortauthors}{S. Chintalapati, J. Bragg, and L.L. Wang}

\begin{abstract}
Figures in scientific publications contain important information and results, and alt text is needed for blind and low vision readers to engage with their content. We conduct a study to characterize the semantic content of alt text in HCI publications based on a framework introduced by \citet{Lundgard2022AccessibleVV}. Our study focuses on alt text for graphs, charts, and plots extracted from HCI and accessibility publications; we focus on these communities due to the lack of alt text in papers published outside of these disciplines. We find that the capacity of author-written alt text to fulfill blind and low vision user needs is mixed; for example, only \percleveltwo of alt texts in our sample contain information about extrema or outliers, and only \perclevelthree contain information about major trends or comparisons conveyed by the graph. We release our collected dataset of author-written alt text, and outline possible ways that it can be used to develop tools and models to assist future authors in writing better alt text. Based on our findings, we also discuss recommendations that can be acted upon by publishers and authors to encourage inclusion of more types of semantic content in alt text. 

\end{abstract}

\ccsdesc[500]{Human-centered computing~Empirical studies in accessibility}

\keywords{accessibility, scientific documents, alt text, dataset}

\maketitle

\section{Introduction}
\label{sec:introduction}

Alternative text (or ``alt text'')
describes the content of a visual graphic or image to those who cannot see them. As such, alt text is an important component of accessible design. 
Most scientific documents use graphics to communicate information alongside text; scientific documents can be especially difficult to make accessible to BLV readers \citep{Bigham2016AnUT}, with a large majority of these paper PDFs lacking usable alt text \citep{Wang2021ImprovingTA}. \citet{Mack2021DesigningTF} conducted a study of BLV users and what they need from alt text, and found that graphs and charts are of special importance to these users, as they can be especially important for conveying results. This, coupled with the finding that the vast majority of scientific figures lack alt text altogether, suggests that even if the rest of the text in a scientific paper were accessible to a BLV reader, that a significant portion of the informational content of these works (figures) remain inaccessible, which can negatively impact reader experience.

Though there have been progress and attempts in automatically generating alt text descriptions of images on web and social media platforms \citep{Wu2017AutomaticAC, Morris2018RichRO, Gleason2020TwitterAA}, these methods do not apply as well to scientific images. From a machine learning perspective, much of the advancement in image recognition and scene understanding in recent years have derived from training neural models on large-scale labeled image datasets (such as ImageNet \citep{Russakovsky2015ImageNetLS} and Google Open Images \citep{Kuznetsova2020TheOI}), datasets that are primarily composed of natural images, which represent only a small proportion of the types of images found in scientific publications. Figures from scientific papers run the gamut of image types, including but not limited to natural images, medical images, diagrams, schematics, a wide array of graphs and charts, as well as combinations of these types, e.g., a medical image annotated with a histogram. Correspondingly, many established image understanding models cannot be directly or easily adapted for the scientific domain. Hybrid crowdsourcing solutions that integrate human experience with machine functionality may offer useful alternative solutions \citep{Gleason2020TwitterAA, Salisbury2017TowardSS, Gurari2018VizWizGC, qian-formative}. For example, \citet{qian-formative} advocate for a hybrid approach to image captioning where machines generate caption units and humans perform stitching. To support these solutions for scientific alt text, we need to better understand the current status of alt text content, and develop tools that can support authors in writing more useful descriptions of scientific figures. 

Several prior studies have attempted to quantify the availability of alt text in scientific documents \citep{Brady2015CreatingAP, Lazar2017MakingTF, Wang2021ImprovingTA}, though none have investigated the content of author-written alt texts and whether they convey adequate information about figures to blind and low vision (BLV) readers. 
In this work, we extract and analyze the content of author-written alt text from papers published by the accessibility and HCI communities, and provide recommendations toward encouraging the inclusion of more types of descriptive information in alt text that may be useful to BLV readers. By extracting realistic author-written alt text, we also provide a useful data resource that can be used to support authors in writing better alt text and study how image understanding models and crowd-authoring techniques can be adapted to more effectively produce figure alt text in the scientific domain.

We process and extract author-written alt text from over 25K publications in the domains of accessibility and HCI, identifying nearly 3.4K pieces of valid alt text from 899 papers. To assess the type of information contained in these alt texts, we use the framework introduced by \citet{Lundgard2022AccessibleVV}, which accounts for four different levels of semantic content that may be conveyed by graphical data visualizations. We assess the semantic content present in the alt text corresponding to figures of graphs, charts, and plots (data visualizations), images that are prevalent in scientific papers, and for which the alt text content can be suitably represented using the levels introduced in the \citet{Lundgard2022AccessibleVV} framework.
We find that though most alt text contain basic information about the graph type, axes labels, and what is plotted, far fewer contain information beyond this.
For example, only \percleveltwo of alt text in our sample discuss extrema or outliers in the data, and only \perclevelthree discuss trends or comparisons. The lack of this type of semantic content in alt text can make it difficult for a BLV user to understand these kinds of images in the way they desire, as found by \citet{Lundgard2022AccessibleVV}.

Our contributions in this work can therefore be summarized as:
\begin{itemize}
    \item An assessment of the semantic information conveyed by author-written alt text of graph and chart figures extracted from papers published in venues representing work in accessibility, HCI, and related areas. We found that levels of covered content are inadequate, even at accessibility and HCI conferences, which have alt text requirements and writing guidelines.
    \item A dataset of \numvalidalttext author-written alt text from HCI publications, of which \totalalttext have been annotated with semantic levels.\footnote{The dataset and annotations are available at \githublink.} The methods used to construct this dataset can be extended to study trends in scientific figure alt text more broadly,
    and our dataset can be used develop tools and models to support alt text authoring. For example, we experiment with training a classifier that identifies semantic levels in text, which could be used to provide feedback to authors as they are writing alt text. We discuss and explore additional opportunities in Section~\ref{sec:uses}.
\end{itemize}

\section{Related Work}
\label{sec:related}
 
We briefly discuss related work on how to write useful scientific alt text (Section \ref{sec:rw_guidelines}), resources and methods for scientific figure understanding (Section \ref{sec:rw_figure_understanding}), resources and methods for automatic alt text generation (Section \ref{sec:rw_alttext_generation}), and other methods for improving figure accessibility (Section \ref{sec:rw_other}).

\begin{table*}[tb!]
\begin{tabular}{L{38mm}L{25mm}L{22mm}L{27mm}l}
    \toprule
    \textbf{Dataset} & \textbf{Domain} & \textbf{Realistic/ Synthetic} & \textbf{Task} & \textbf{Size of dataset} \\
    \midrule

    FigureSEER \cite{Siegel2016FigureSeerPR} & Scientific figures & Realistic & Image classification (6 classes) & 60K figures \\
    DocFigure \cite{Jobin2019DocFigureAD} & Scientific figures & Realistic & Image classification (28 classes) & 33K images \\
    SlideImages \cite{Morris2020SlideImagesAD} & Educational illustrations & Realistic & Image classification (8 classes) & 3K images \\
    \midrule

    FigureQA \cite{Kahou2018FigureQAAA} & Scientific graphs and charts & Synthetic & VQA & 100K images \\
    DVQA \cite{Kafle2018DVQAUD} & Bar charts & Synthetic & VQA & 300K images \\
    PlotQA \citep{Methani2020PlotQARO} & Plots & Synthetic plots; Realistic data & VQA & 224K images \\
    \midrule

    SciCap \cite{Hsu2021SciCapGC} & Scientific figures & Realistic & Image captioning & 416K figures \\
    ImageClefMed Caption \cite{Nicolson2021AEHRCCA} & Medical images & Realistic & Image captioning & 5K images \\
    \midrule

    FigCAP \cite{Chen2019NeuralCG, Chen2019FigureCW} & Bar charts, pie charts, line plots & Synthetic & alt text generation & 110K figures \\
    \bottomrule
\end{tabular}
\caption{Datasets for scientific figure understanding}
\label{tab:datasets}
\end{table*}

\subsection{Guidelines for writing scientific alt text}
\label{sec:rw_guidelines}

The Web Content Accessibility Guidelines (WCAG) \citep{Chisholm2001WebCA, Caldwell2008WebCA} contain guidance on when alt text should be provided and the suggested content for the alt text. The National Center for Accessible Media (NCAM) has published guidelines including high-level recommendations for writing alt text for graphs, suggesting that a complete description should include text describing (i) the layout of the graph, (ii) the location of variables on the graph, and (iii) for static graphs, the overall trends presented, and for dynamic graphs, summary information such as the range of the axes.\footnote{\href{https://www.wgbh.org/foundation/ncam/guidelines/accessible-digital-media-guidelines}{https://www.wgbh.org/foundation/ncam/guidelines/accessible-digital-media-guidelines}} The Benetech Diagram Center also provides image description guidelines with the goal of making it easier, cheaper, and faster to create and use accessible digital images.\footnote{\href{http://diagramcenter.org/making-images-accessible.html}{http://diagramcenter.org/making-images-accessible.html}} The referenced documentation includes 
both general best practices concerning aspects such as style and language that apply to every type of image, along with specific considerations for bar graphs, pie graphs, line graphs, and scatter plots such as listing the numbers in a pie graph from smallest to largest, and focusing on the change of concentration in scatter plots. Several academic publishers have also provided research-based guidelines for improving the accessibility of digital media. For example, the Association for Computing Machinary (ACM) strongly encourages authors to provide alt text for images and charts, and includes instructions for authors such as not duplicating the caption text and providing keywords.\footnote{\href{https://authors.acm.org/proceedings/production-information/describing-figures}{https://authors.acm.org/proceedings/production-information/describing-figures}}

\citet{Lundgard2022AccessibleVV} introduced a four-level conceptual model describing the semantic content of information that should be present in alt text descriptions of scientific data visualizations. Level 1 includes construction details such as the type of figure (e.g., bar plot or line plot), and labels of the axes. Level 2 includes statistics about the figure data, such as extremes and correlations. Level 3 includes larger takeaways, such as trends and patterns in the data. Level 4 includes domain-specific insights and societal context for the figure data. The authors conducted studies including BLV users, and found that these users gained the most information from textual descriptions conveying information from semantic levels 1--3 \citep{Lundgard2022AccessibleVV}. This finding corresponds to the recommendations of alt text content made by the NCAM. Though this framework is not a guideline document \emph{per se}, we adopt it for this study in order to evaluate the quality of author-written alt text for graphs and charts found in scientific publications. It has been validated through studies with BLV participants, and we are unaware of alternative frameworks.

Towards alt text preferences, \citet{bennett-chi21} 
conducted a study on best practices for describing race, gender, and disability status in alt text, and found that people in photographs preferred to be described with the language that they use to talk about themselves, and that descriptions of concrete visual details were more appropriate than language around identities. Though this does not apply directly to scientific figures, alt text written by someone other than the authors of a paper may want to consider how the authors intended for the figure to be understood as well as the language used by the authors in the rest of the publication.
After reviewing BLV people's experiences with digital image types such as news articles and employment websites, \citet{stangl-what-want} found that a one-size-fits-all approach for image descriptions is not ideal. Similarly, through interviews with screen reader users, \citet{Mack2021DesigningTF} found that different BLV users had varying preferences about the level of detail that they found to be most helpful in alt text, although most users concluded that both brevity and the availability of detailed information were desirable traits. Such findings should be kept in mind when authoring effective alt text.

\subsection{Interfaces for authoring alt text}
\label{sec:interfaces}

\citet{Morash2015GuidingNW} developed interfaces to guide novice web workers in writing descriptions of scientific images. The authors queried workers for information about select image attributes based on the NCAM guidelines, on attributes such as image type, title, and units shown. They found that the templated query method was preferred by the workers and produced better image descriptions.

\citet{Mack2021DesigningTF} built a prototype interface for authoring alt text, and measured the quality of alt text on a four-point scale based on three interface variations: the current PowerPoint interface; a free-form interface, where suggestions were presented as a bulleted list; and a template interface, where each prompt was listed separately and included a text box to respond to that prompt. Participants who use screen readers were asked to rank the quality of alt text written under the PowerPoint, free-form, and template interfaces, they found that, in general, the free-form interface encouraged authors to write alt text that is more closely aligned with the preferences of screen reader users. 

\subsection{Automated methods for scientific figure understanding}
\label{sec:rw_figure_understanding}

Scientific figure understanding tasks such as figure classification, visual question-answering (VQA), or image captioning have received significant attention from the AI community in recent years \citep{Siegel2016FigureSeerPR, Jobin2019DocFigureAD, Morris2020SlideImagesAD, Kahou2018FigureQAAA, Kafle2018DVQAUD, Methani2020PlotQARO, Levy2021ClassificationRegressionFC, Chen2019NeuralCG, Chen2019FigureCW, Hsu2021SciCapGC, Nicolson2021AEHRCCA, Qian2021GeneratingAC}. In Table~\ref{tab:datasets}, we describe a number of datasets that have been introduced to train models and evaluate their performance on these tasks.

Datasets introduced for scientific image classification include FigureSeer \citep{Siegel2016FigureSeerPR}, DocFigure \citep{Jobin2019DocFigureAD}, and SlideImages \citep{Morris2020SlideImagesAD}. All three datasets include realistic images extracted from scientific papers, along with labels to classes such as graph, medical image, or natural image. These datasets have been used to train models that can detect figure type, which is an essential piece of information that should be available in alt text. However, figure type is only one of many pieces of information that BLV users may need to understand the content of an image, and therefore these datasets are of limited use in the alt text generation setting.

Towards more detailed figure understanding, datasets such as FigureQA \citep{Kahou2018FigureQAAA}, DVQA \citep{Kafle2018DVQAUD}, and PlotQA \citep{Methani2020PlotQARO} have been introduced. These datasets are made up of synthetically generated graphs and charts, along with associated questions and answers about the graph contents. For example, questions may be related to the graph title, axes, the $x$ and $y$-axes values associated with specific data series, or the names of each data series. In PlotQA \citep{Kafle2018DVQAUD}, many questions also go beyond the structure of the plot and may require data retrieval or additional reasoning (e.g. What is the average difference between two data series?) Models trained on these datasets have shown improving performance \citep{Kahou2018FigureQAAA, Levy2021ClassificationRegressionFC}, though because the plots in these datasets are generated synthetically, they may not transfer well to graphs found in actual scientific publications, which are significantly more noisy, diverse, and variable than those found in these datasets. Also, though figure understanding through VQA is related to the task of producing alt text, the task of VQA itself does not produce a coherent textual description of the figure, which is the desired outcome for alt text. The FigCAP project \citep{Chen2019NeuralCG, Chen2019FigureCW} attempts to bridge this divide by deriving figure descriptive text from the questions and answers of the FigureQA \citep{Kahou2018FigureQAAA} dataset; though FigCAP refers to itself as a figure captioning dataset, the ``captions'' they provide are more similar to the notion of alt text, including descriptions about the figure structure, axes labels, data values, etc.

Recent work has aimed to automatically generate captions for both natural images \citep{Chen2015MicrosoftCC, Sharma2018ConceptualCA} and figures \citep{Hsu2021SciCapGC, Nicolson2021AEHRCCA}; however, caption generation and alt text generation are not the same task, and could be said to have differing goals. Captions are intended to be consumed by all readers, and contain information that complements the content of the image; while alt text is meant to explain the informational content of the image for users who cannot see it. SciCap \citep{Hsu2021SciCapGC} introduces a large dataset of graph figures and captions derived from arXiv; captions are those originally written by authors, and are post-processed to remove tokens corresponding to numbers and equations. The ImageClefMed Captioning task released a relatively much smaller dataset focused on captioning of medical images derived from scientific publications \cite{Nicolson2021AEHRCCA}. 

As far as we know, there are no datasets available for studying scientific figure alt text generation with realistic, author-written alt text. Though FigCAP \citep{Chen2019NeuralCG, Chen2019FigureCW} and FigJAM \cite{Qian2021GeneratingAC} explore text generation in the alt text setting, the images and target texts used are synthetic (derived from FigureQA \citep{Kahou2018FigureQAAA}) and not representative of the analogous task in a realistic setting.

\subsection{Automated methods for alt text generation}
\label{sec:rw_alttext_generation}

Office 365 generates alt text for any image or figure pasted into Microsoft PowerPoint.\footnote{\href{https://www.microsoft.com/en-us/microsoft-365/powerpoint}{https://www.microsoft.com/en-us/microsoft-365/powerpoint}} Though easy to use, the feature shows limited performance on scientific figures, usually only describing the type of the figure. For example, for the figures in Table~\ref{tab:example_figures}, the corresponding alt texts generated by Office 365 are, respectively: ``Chart, bar chart,'' ``Chart, line chart,'' and ``Chart.'' Though Office 365 usually identifies the correct type of chart, no other information about the figure content is generated, and the resulting alt text is of limited use to the reader. 

\citet{Qian2021GeneratingAC} created synthetic datasets for figure alt text generation by adapting the FigureQA \citep{Kahou2018FigureQAAA} and DVQA \citep{Kafle2018DVQAUD} datasets for figure VQA. Figures in both datasets are synthetic (not from scientific papers), and alt text units are derived from the data and information used to construct each figure. The authors then trained the FigJAM model, which generates alt text descriptions based on the multimodal inputs of the raw figure image and figure metadata \citep{Qian2021GeneratingAC}. Though performance on synthetic data was shown to be good, the model may not generalize to realistic figures found in scientific papers, which exhibit significantly more variability than those evaluated by \citet{Qian2021GeneratingAC}.\footnote{The authors do not release their dataset or pretrained models, so we are unable to perform a comparative analysis on realistic figures derived from scientific papers.}

Researchers have developed tools for generating alt text information for images on non-scientific social platforms.
Gleason et al. \cite{Gleason2020TwitterAA} addressed the accessibility barrier on Twitter by creating a browser extension which adds alt text to Twitter using six methods, such as reverse image searching and automatic image captioning. However, this project focused on Twitter images, which differ from realistic scientific figures. Additionally, Wu et al. \cite{Wu2017AutomaticAC} deployed an automatic alt text system to identify faces, objects, and themes for photos on Facebook in order to make them more accessible to screen reader users. The domain of natural images on social media again significantly differs from our domain of scientific graphs and charts.

\subsection{Other methods for improving figure accessibility}
\label{sec:rw_other}

Researchers have also developed alternatives to alt text for improving figure accessibility. ChartSense \cite{Jung2017ChartSenseID} and PlotDigitizer\footnote{\href{http://plotdigitizer.sourceforge.net/}{http://plotdigitizer.sourceforge.net/}} are chart data extraction methods which convert chart images into structured data tables. However, charts of the same type are too diverse in style to apply a single extraction algorithm, algorithms have trouble interpreting overlapped visual entities, and there is no text-region detection algorithm for chart images with sufficient accuracy \cite{Jung2017ChartSenseID}. Auditory graphs, tactile graphs, and various multimedia approaches have also been introduced to improve the accessibility of graphs and charts \cite{Engel2017ImproveTA, Nazemi2013AMT}.

Crowdsourcing, in which a group of non-experts completes a task that is currently infeasible to accomplish via automated methods, has also been used to good effect for making images more accessible. \citet{Salisbury2017TowardSS} explored a novel approach in which crowdworkers were paired together to create image descriptions. Crowdworkers were asked questions to extract desired details about an image such as the location of the image and what emotions the image evoked. Platforms such as VizWiz \citep{Bigham2010VizWizNR} or Be My Eyes\footnote{\href{https://www.bemyeyes.com/}{https://www.bemyeyes.com/}} connect BLV users to sighted crowdworkers and volunteers via an app for assistance with questions and daily tasks. Building upon the successes of the VizWiz Grand Challenge \citep{Gurari2018VizWizGC}, a similar solution could be created to connect BLV researchers with volunteers that can provide suitable descriptions or answers to questions about scientific figures. \citet{Mack2021DesigningTF} built user interfaces for both authoring alt text and providing feedback on automatic alt text, which could be theoretically connected to a crowdsourcing platform and adapted to collect alt text for scientific figures based on user demand.

Lastly, researchers have developed methods for surfacing alt text that do not involve manual effort from humans or automated alt text generation. \citet{Guinness2018CaptionCE} found that many images appear in several places across the web, and used this insight to develop Caption Crawler. This system uses reverse image search to find alt text from similar images available on the web and surfaces these alternate descriptions to the user. This method works quite well in the domain of natural images, where many similar images of places or things can be found on the open internet.

\section{Methods}
\label{sec:methods}

We extract and analyze the presence and content of alt text for graphs, charts, and plots in scientific papers. These figures are typically used to visualize data and results, are of special importance to BLV readers \citep{Mack2021DesigningTF}, and are the types of images for which \citet{Lundgard2022AccessibleVV} have defined semantic levels. The \citet{Lundgard2022AccessibleVV} framework describes four levels of semantic content, organized by increasing complexity: 

\begin{itemize}
    \item Level 1: enumerating visualization construction details (e.g., type, marks, and encodings) 
    \item Level 2: identifying statistical concepts and relations (e.g., extremes and correlations) 
    \item Level 3: characterizing perceptual and cognitive phenomena (e.g., trends and patterns)
    \item Level 4: articulating domain-specific insights or societal context. 
\end{itemize}

\noindent \citet{Lundgard2022AccessibleVV} found that levels 1--3 were reported most useful by blind and low vision readers. Level 4, which incorporates significantly more subjective information, was found to be less essential to understanding; in fact, a majority of blind readers in their study (63\%, n=19) believed that figure alt text should not contain level 4 content.

\subsection{Sampling papers and extracting author-written alt text}

Our goal is to construct a dataset of author-written alt text by automatically sampling and extracting alt text from papers. We start with the set of papers from two conferences: ACM CHI Conference on Human Factors in Computing Systems (CHI) and ACM SIGACCESS Conference on Computers and Accessibility (ASSETS) published in the years 2010-2020. We identify these papers using the ACM's reported DOIs, and link these to PDFs in the Semantic Scholar corpus \citep{Ammar2018ConstructionOT}. This yields 5218 PDFs. We then iteratively extend our paper sample to include all papers written by authors who have published in CHI and/or ASSETS. We prioritize authors based on the frequency of their publications in CHI and ASSETS, and retrieve these authors' other publications using the Semantic Scholar API.\footnote{\href{https://api.semanticscholar.org/}{https://api.semanticscholar.org/}} From these queries, we assemble a further sample of 20000 paper PDFs to process and extract alt text.

CHI and ASSETS are premiere conferences on human-computer interaction and accessible computing, and both conferences have a history of soliciting accessible paper submissions\footnote{CHI directs authors to the SIGCHI guidelines for an accessible submission: \href{https://sigchi.org/conferences/author-resources/accessibility-guide/}{https://sigchi.org/conferences/author-resources/accessibility-guide/} and ASSETS provides these instructions: \href{https://assets21.sigaccess.org/creating_accessible_pdfs.html}{https://assets21.sigaccess.org/creating\_accessible\_pdfs.html}} and requiring authors to provide alt text for figures. We therefore expect a much higher percentage of papers published at these venues to contain valid alt text.
Similarly, our rationale for sampling additional papers published at other venues by authors who have published at CHI and ASSETS is based on our hypothesis that such authors are more likely to write alt text in general, even if another publishing venue may not require alt text for submission. Initially, we intended to extract alt text from a stratified random sample of papers representing all fields of study, but a pilot attempt showed that random sampling would yield virtually no alt text. We arrived at this conclusion after processing a stratified sample of 5000 PDFs from 2010 to 2020 and extracting only a single piece of descriptive alt text. Given the time and expense of processing PDFs to extract alt text at scale, we decided that the stratified random sample strategy was untenable.

For each paper in our sample, we followed a three step process to extract alt text. 
First, we processed the PDF using Adobe Acrobat Pro\footnote{\href{https://www.adobe.com/acrobat/acrobat-pro.html}{https://www.adobe.com/acrobat/acrobat-pro.html}} to convert the PDF to HTML using the Adobe Acrobat Pro Action Wizard. We used Adobe Acrobat Pro rather than an alternate programmatic approach because we were unable to identify a PDF processing library capable of extracting alt text.\footnote{We experiment with several other widely available PDF libraries and conversion tools, including PDFTOHTML (\href{http://pdftohtml.sourceforge.net/}{http://pdftohtml.sourceforge.net/}), PDFMiner (\href{https://github.com/euske/pdfminer}{https://github.com/euske/pdfminer}), pdf2xml (\href{https://sourceforge.net/projects/pdf2xml/}{https://sourceforge.net/projects/pdf2xml/}) etc., and found that none of these tools allowed access to the embedded alt text. Scaling in Adobe Acrobat Pro was the only working solution we were able to identify.} Second, we extracted alt text from the converted HTML document. Third, we filtered the extracted alt text according to a set of heuristics. Our pilot attempts at extracting alt text revealed that most of the extracted alt text consisted of uninformative short descriptions like ``Image'' or file paths like ``C:\textbackslash\textbackslash path\_to\textbackslash figure1.jpeg,'' so we defined a set of filtering criteria to remove these. 
We also filtered out alt text shorter than 80 characters since many shorter alt text fall under the category of uninformative short descriptions. To determine this 80-character threshold, we analyzed a sample of 100 extracted alt text to find a limit that maximizes recall without sacrificing precision. We refer to the alt text that pass these filtering criteria as ``valid alt text.''

For all valid alt texts, we identify those that are likely to correspond to graphs, charts, and plots. We iteratively defined another set of heuristics: a list of words and phrases that correspond to graphs and charts (e.g., ``graph'', ``chart'', ``error bar''; the full list of terms are provided in Appendix~\ref{app:heuristic_terms}). For each figure, we search for token matches in the alt text and image caption against this list of terms, and retain only figures and alt text matching at least one term. The alt text of the matching figures are then annotated with the \citet{Lundgard2022AccessibleVV} semantic levels.

\begin{table*}[tbhp!]
    \centering
    \begin{tabular}{L{60mm}L{95mm}}
        \toprule
        Figure \& Source & Author-written alt text with annotations \\
        \midrule
        {\raisebox{-\totalheight}{\includegraphics[width=50mm]{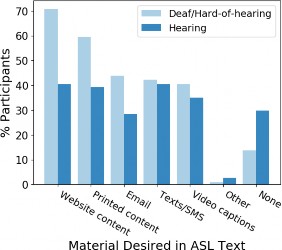}}} \vspace{2mm} \begin{center}Figure 4 reproduced from \citet{Bragg2018DesigningAA}\end{center} &  ``Figure 4: Materials participants reported wanting to read in ASL text \blue{(L1)}. This figure presents a bar chart, with separate bars for DHH (light blue) and hearing (dark blue) populations \blue{(L1)}. Y-axis is \% participants, ranging from 0-70 \blue{(L1)}. X-axis is Material Desired in ASL Text \blue{(L1)}. sorted by DHH popularity (most popular first): Website content, Printed content, Email, Texts/SMS, Video captions, Other, None \blue{(L1)}.'' \\
         & \\
        {\raisebox{-\totalheight}{\includegraphics[width=57mm]{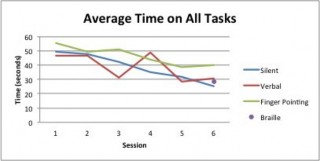}}} \vspace{2mm}
        \begin{center}Figure 4 reproduced from \citet{Baker2014TactileGW}\end{center} & ``A line chart showing the average time it took participants on all tasks \blue{(L1)}. The y-axis of the chart is time in seconds (ranges from 0 to 60), the x-axis of the chart is session number (ranges from 1 to 6) \blue{(L1)}. There is a line for the three modes: Silent, Verbal and Finger Pointing \blue{(L1)}. They all appear to be going down, but there is a big spike in the Verbal mode line at session 4 \blue{(L3)}. In general, the Finger Pointing mode is the highest (takes the most time), the Silent mode is next and the Verbal takes the least amount of time, although in the fourth and sixth sessions, the Verbal line is above the Silent one \blue{(L3)}. There is a dot corresponding to the Braille mode at Session 6, it is between the Verbal and Silent modes \blue{(L1)}.'' \\
         & \\
        {\raisebox{-\totalheight}{\includegraphics[width=57mm]{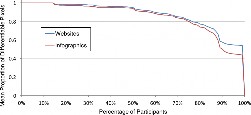}}} \vspace{2mm}
        \begin{center}(Figure 7 reproduced from \citet{Reinecke2016EnablingDT})\end{center} &  ``Plot of mean proportion of image pixels differentiable (independent) for 0\% - 100\% of the population (dependent) for websites and infographics \blue{(L1)}. Increasing from 0\% of the population, both plots start at 100\% differentiable and gradually fall to 80\% differentiable at 75\% of the population \blue{(L2, L3)}. Plots begin to diverge here as they both fall off more quickly until a discontinuity plateau is reached at 88\% of the population (websites = 60\% differentiable, infographics = 50\% differentiable) \blue{(L2, L3)}. Plateau gradually declines to 99\% of population (websites = 55\% differentiable, infographics = 45\% differentiable), and then both plots fall to 0\% differentiable for 100\% of the population \blue{(L2, L3)}.'' \\
         & \\
        \bottomrule
    \end{tabular}
    \caption{Example figures and author-written alt text with annotated semantic levels added to the end of each sentence, with the prefix L, in parentheses and colored \blue{(blue)}.}
    \label{tab:example_figures}
\end{table*}

\subsection{Annotation of alt text semantic levels}

To study what types of content are present in author-written alt text, we ask annotators to assess the semantic content levels present in each sentence of each piece of alt text. We split alt text into sentences using the scispaCy NLP library \citep{Neumann2019ScispaCyFA}. Annotators are shown these sentences along with the corresponding figure caption and a link to view the figure.

Six label options were provided for each sentence:

\begin{itemize}
    \item Level 1: Figure logistics

    \item Level 2: Statistical properties and comparisons
    \item Level 3: Complex trends and patterns in data
    \item Level 4: Domain-specific insights or societal concepts to help explain Level 3 trends
    \item This alt text contains no levels of content
    \item This image is not a graph or chart
\end{itemize}

\noindent If a figure is not a graph or chart, the annotator is instructed to select the last option. Otherwise, up to three semantic levels could be selected for each sentence. These label options were adapted from the level descriptions given by \citet{Lundgard2022AccessibleVV} and shortened and simplified to make them easier for annotators to understand. Annotators were given an additional instruction document that provides more detail on each of the label options, along with examples of sentences corresponding to each label option.

The annotators were instructed to exhaustively label each sentence with all of the levels it contains. We recruited two annotators through the UpWork platform.\footnote{\href{https://www.upwork.com/}{https://www.upwork.com/}} The annotators had undergraduate-level education in math, statistics, and materials science, and had previous experience reading graphs, plots, and other scientific figures. The alt text retained after the filtering steps described in the previous section were split into sets of 100 alt texts each for annotation. Two individuals annotated an initial sample of 100 alt texts (298 sentences) to refine the task and ensure high annotator agreement. The inter-annotator agreement computed over this sample was 87.6\%, with $\alpha$ = 0.80, indicating very good agreement. We further clarified the instructions following a discussion of disagreements. Given the high agreement level, a single annotator annotated the remaining alt text. For the final analysis, the first annotator's annotations are used for the sample that was doubly annotated. Examples of extracted alt text and the corresponding annotated semantic levels are provided in Table~\ref{tab:example_figures}.

\section{Results}
\label{sec:results}

We analyze the semantic content of the extracted alt text, and attempt to answer the following research questions:

\begin{itemize}
    \item \textbf{RQ1: What is the distribution of semantic content in author-written alt text?}

    We want to determine the proportion of alt text containing level 1, 2, 3, and 4 content. Of these, what proportion contains levels 1--3 content, which satisfies most BLV user needs? Correspondingly, which semantic levels are most often missing?
 
    \item \textbf{RQ2: How does the distribution of semantic content in alt text change over time?} We want to determine whether the presence of levels 1-3 semantic content is increasing over time, and by how much. We expect that with improvements in alt text awareness and workflows over time, that the amount of content available in alt text should correspondingly increase.
    \item \textbf{RQ3: How does length of alt text correlate with semantic levels?} 
 
    \citet{Mack2021DesigningTF} find that there is tension between detail and brevity in alt text. The ideal alt text may vary based on user needs, but should balance length and completeness of semantic content. We determine the relationship between length and presence of semantic levels using our data.

\end{itemize}

\subsection{Descriptive statistics}

We process \numtotalpdfs total paper PDFs to extract alt text. Of these, \numconvertedpdfs (\percconvertedpdfs) are successfully converted to HTML by Adobe Acrobat Pro. Only \numpaperswithvalidalttext (\percpaperswithvalid) of these converted documents contain at least one piece of valid alt text. Around 2048 pieces of alt text corresponded to file paths and 2545 alt texts did not meet our length criteria, and were filtered out; all other alt texts that were filtered out did not have any content besides ``image.'' After this filtering, the \numpaperswithvalidalttext papers contain \numvalidalttext valid author-written alt texts. Using our keyword heuristics, we determine that \numannotated of these alt texts are likely to correspond to graphs or charts. Based on a cursory examination, the alt texts that were filtered out using these keyword heuristics consist primarily of natural images and diagrams. 

We ask annotators to assess the semantic levels present in each sentence of the remaining \numannotated alt text. Of these, \totalalttext figure alt texts (consisting of \totalalttextsents sentences) are labeled as belonging to graphs, charts, and plots by our annotators, indicating that our keyword heuristics have approximately 50\% precision. Alt texts of figures corresponding to these data visualizations are further annotated for semantic content. Several examples of author-written alt text and semantic level annotations are given in Table~\ref{tab:example_figures}. In Table~\ref{tab:filtering}, we provide the numbers of PDFs processed and alt text retained after each filtering step.

\begin{figure}[t!]
    \centering
    \includegraphics[width=0.49\textwidth]{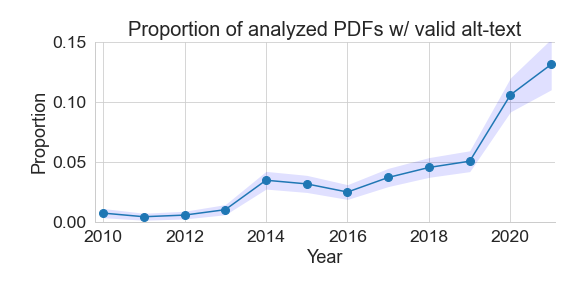}
    \caption{The proportion of PDFs in our analyzed sample which contain valid alt text over time, with 95\% confidence intervals computed through bootstrap resampling.}
    \label{fig:alttext_over_time}
    \Description{This is a line chart entitled "Proportion of analyzed PDFs with valid alt text" that plots the proportion of PDFs with valid alt-text by year. The years are plotted on the horizontal x-axis from 2010 to 2021 with an increment of 1 year. The proportions of analyzed PDFs with valid alt-text are plotted on the vertical y-axis from 0.00 to 0.15 with an increment of 0.05. The year 2011 has the lowest proportion of valid alt text over time. The year 2021 has the highest proportion of valid alt text over time. Alt text proportion increases over time, with slight fluctuations. Years 2014 to 2016 have the most noticeable drop in alt text proportion, but that drop is only a decrease by about 1\%, from 3.5\% to 2.5\%. The greatest increase happens from 2019 to 2021, with 2019 having a proportion of 5\% and 2021 having a proportion of approximately 13\%. There is also a larger than average increase from 2013 to 2014, with valid alt text proportion increasing from approximately 1\% to 3.5\%.}
\end{figure}

Figure~\ref{fig:alttext_over_time} shows the proportion of PDFs in our sample that contain any valid alt text, and how this proportion changes in our sample over the last decade. There is a slight increase in alt text coverage in 2014; this is the same year that CHI specified that alt text is required in submissions. We also observe that the proportion of papers with valid alt text has improved over time, especially in the past few years, although the overall proportion is still quite low (below 15\%). We note that Figure~\ref{fig:alttext_over_time} does not indicate the actual proportions of papers from these years that have valid alt text (we do not process the version of record for all papers in our sample due to the difficulty in ascertaining these versions and copyright challenges in obtaining them; we also cannot guarantee that the pipeline we use succeeds in extracting alt text for all of the papers we process). Rather, the figure describes our success rate in creating this dataset, and provides some sense of the trend towards more valid alt text in recent years. 

\begin{table}[t!]
\begin{tabular}{L{62mm}l}
    \toprule
    Processing step & Count \\
    \midrule
    Total PDFs processed & \numtotalpdfs \\
    PDFs successfully converted to HTML & \numconvertedpdfs (\percconvertedpdfs) \\ 
    Papers with at least one valid alt text & \numpaperswithvalidalttext (\percpaperswithvalid) \\ 
    Pieces of valid alt text & \numvalidalttext \\ 
    Number of figure alt text annotated (heuristically filtered to likely be graphs or charts) & \numannotated \\ 
    Number of annotated figure alt text that correspond to graphs or charts & \totalalttext \\
    \bottomrule
\end{tabular}
\caption{The numbers of papers and figure alt text that remain after each filtering step.}
\label{tab:filtering}
\end{table}

\subsection{Analysis of alt text semantic content}

\subsubsection*{Distribution of semantic content in alt text (RQ1)}

\begin{figure}[tb!]
    \centering
    \includegraphics[width=0.45\textwidth]{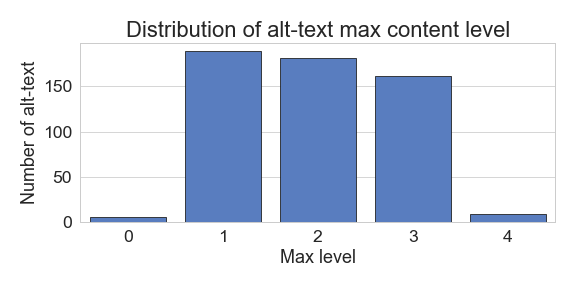}
    \includegraphics[width=0.45\textwidth]{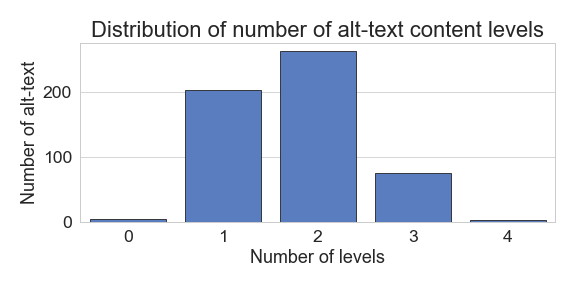}
    \caption{Distribution of the maximum level (top) and total number of levels (bottom) of semantic content found in the sample of annotated author-written alt text.}
    \label{fig:counts}
    \Description{These are two vertical bar charts entitled "Distribution of alt-text max content level" and "Distribution of number of alt-text content levels." The "distribution of alt-text max content level" bar chart plots the max level by the number of alt-text with that max level. Max level is plotted on the horizontal x-axis from 0 to 5, with 0 representing no levels of content. Number of alt-text is plotted on the vertical y-axis from 0 to 200. The highest frequency max level is level 1, with 189 occurrences of alt-text. The lowest frequency max level is level 0, with 5 occurrences of alt-text, followed by level 4, with 9 occurrences of alt-text. Number of alt-texts does not increase as max level increases, since levels 1 and 2 have the largest number of alt-text, with level 3 having slightly less and level 4 having drastically less. The "Distribution of number of alt-text content levels" bar chart plots the number of levels by the number of alt-text in each of those categories. Number of levels is plotted on the horizontal x-axis from 0 to 4, with 0 representing no levels of content and 4 representing 4 levels of content. Number of alt-text is plotted on the vertical y-axis from 0 to 300. The highest frequency number of levels is 2 levels, with 262 corresponding pieces of alt-text. The lowest frequency number of levels is 4 levels, with 3 corresponding pieces of alt-text. The number of alt-texts corresponding to the number of levels increases then decreases, with it increasing from 0 levels to 1 level and from 1 level to 2 levels, and then decreasing from 2 levels to 3 levels and 3 levels to 4 levels.}
\end{figure}

In Figure~\ref{fig:counts} (left), we present the maximum level of content found in each figure alt text in our sample. We observe a fairly evenly distribution between max levels 1, 2, and 3. Recall that in \citet{Lundgard2022AccessibleVV}, BLV users found levels 1--3 content to be most useful. Although some figures in our sample have alt text with level 3 content, over two-thirds of the figure alt texts that we examined do not contain level 3 content (level 3 content describes trends and patterns). At the same time, over one-third of the figure alt texts lack both level 2 and level 3 content, meaning that there is no information on extrema and outliers in addition to trends and patterns. The lack of such content can make it more difficult for BLV users to acquire the information that they need from these graphs and charts.

In Figure~\ref{fig:counts} (right), we show the total number of levels of content present in all figure alt text we analyzed. We find that the vast majority of alt text in our sample only contain one or two levels. This means that though the maximum levels of content are somewhat evenly distributed between levels 1, 2, and 3, only one or two of these levels are present in most figure alt texts. 

\subsubsection*{Semantic content in alt text over time (RQ2)}

\begin{figure}[tb!]
    \centering
    \includegraphics[width=0.49\textwidth]{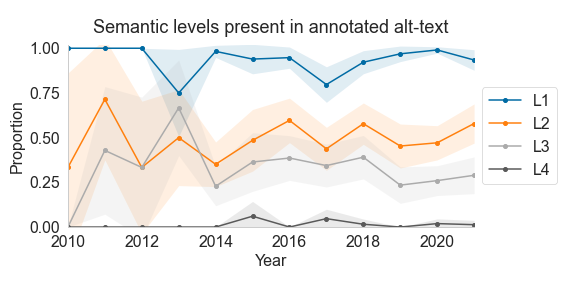}
    \caption{Proportion of alt text containing text of each semantic level over time with 95\% confidence intervals computed using bootstrap resampling.}
    \label{fig:levels_over_time}
    \Description{This is a multi-line chart entitled "Semantic levels present in annotated alt-text" that plots the proportion of alt-text that contains each level by year. The years are plotted on the horizontal x-axis from 2010 to 2021 with an increment of 1 year. The proportion of alt-text containing text of each semantic level is plotted on the vertical y-axis from 0.0 to 1.00 with increments of 0.25. Level 1 has the highest proportion over time, while level 4 has the lowest proportion over time. The proportion of levels 1 to 4 seem to fluctuate with no clear increase over time. The level 1 proportion fluctuates slightly while remaining near 1.0, with an exception from 2012 to 2013 where there is a decrease in proportion, and 2013 to 2014, where the proportion subsequently increases back to around 1.0. The level 4 proportion remains almost constant at 0.0, with slight fluctuations over time. Levels 2 and 3 show the most increase over time, with noticeable increases from 2010 to 2011, and drops from 2011 to 2021.}
\end{figure}

Figure~\ref{fig:levels_over_time} shows the proportion of alt text from each year that contain text of each of the semantic levels. Though the vast majority of alt text contain level 1 information, a much lower proportion contain level 2 and 3 information. Over time, there have not been significant changes to the proportion of alt text that contains level 2 and 3 information.

\subsubsection*{Relationship between alt text length and content (RQ3)}

\begin{figure}[t!]
    \centering
    \includegraphics[width=0.46\textwidth]{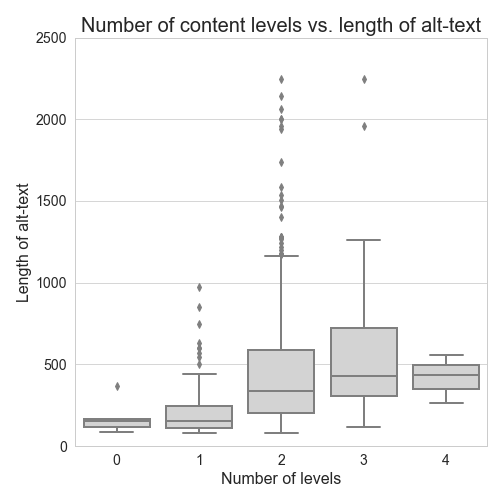}
    \caption{Relationship between length of alt text (character count) and the number of semantic levels of information represented. Though length correlates with the number of levels, there are many longer alt texts that do not necessarily contain more levels of semantic content.}
    \label{fig:alttext_length}
    \Description{This is five side-by-side vertical box plots entitled "Number of content levels vs. length of alt-text" that plots the number of semantic levels of information by the length of alt text, represented by character count. The levels are plotted on the horizontal x-axis from 0 to 4, with 0 representing no levels of content and 4 representing 4 levels of content, and one box plot for each number of levels. The length of alt-text is plotted on the vertical y-axis from 0 to 2500, with increments of 500. On average, the median length of alt-text containing 3 or 4 levels of content is slightly higher than the median length of alt-text containing 0, 1, or 2 levels of content. However, this difference is slight. There is a correlation present between length of alt text and number of levels, but there is overlap between the box plots for 2, 3, and 4 levels of content, meaning that there are longer alt texts that do not contain more semantic content, and there are shorter alt texts that do contain more content.}
\end{figure}

In Figure~\ref{fig:alttext_length}, we show the relationship between the length of alt text and the number of levels of content that it contains. Alt text containing more levels of information tend to be longer, though alt text of comparable length can have different numbers of levels present (as indicated by the overlapping boxes across levels). This suggests that alt text does not have to be longer in order to have more levels of content. In regards to the balance between brevity and detail, authors may want to optimize for the amount of information included in alt text without resorting to writing something overly long.

\section{Dataset uses}
\label{sec:uses}

Alongside our analysis, we release our dataset of \numvalidalttext alt text collected from \numpaperswithvalidalttext HCI
publications.\footnote{Data and analyses are available at \githublink.} Of these, the contents of \totalalttext alt texts (\totalalttextsents sentences) are annotated with the semantic levels introduced by \citep{Lundgard2022AccessibleVV}. This dataset can be used to develop tools to support authors in writing better alt text, or to develop authoring tools or models for producing alt text where none are available. We discuss some of these potential applications below.

\subsection{Improving author-written alt text \& supporting reading interfaces}
\label{sec:classifier}

To improve authoring of alt text, we can develop tools to help identify potentially missing content in alt text and prompt authors to add such information during the authoring and editing process. For graphs and charts, the semantic levels serve as a proxy for content. Based on the findings by \citet{Lundgard2022AccessibleVV}, BLV users found levels 1--3 information most beneficial, and a tool aimed to improve alt text for graphs and charts could assess alt text quality based on the presence or absence of information at these levels. In other words, we could train a classifier based on the semantic level annotations in our dataset to detect which levels are present and build a tool that prompts authors for the missing levels, similar to interfaces that have been built in the past for other tasks like providing peer feedback~\cite{critiquekit}. Such a classifier could also be used to automate trend monitoring for alt text content, enabling expanded and continually updating versions of our analysis with reduced human labor.

To test out the viability of this theory, we train several classification models using our collected annotations. We approach this as a multi-class, multi-label classification problem. The classes are the four semantic levels, and up to four labels can be assigned to a single piece of text. The input to the model is a single sentence of alt text, and the output is a distribution of labels over the four classes. We experiment with a Random Forest classifier using \emph{tf-idf} word representations, as well as classifiers based on BERT~\citep{Devlin2019BERTPO} and SciBERT~\citep{Beltagy2019SciBERTAP}. We use 5-fold cross-validation to train and evaluate all models. The training data is split into folds preserving each alt text as a unit (\totalalttext instances), while text and labels are provided to the model at the sentence level (\totalalttextsents sentences). We report mean accuracy and F1 over all five folds for each model in Table~\ref{tab:classifier}.

\begin{table}[tb!]
    \centering
    \begin{tabular}{L{30mm}cc}
        \toprule
        Model & Accuracy & Micro-F1 \\
        \midrule
        Random Forest (\emph{tf-idf}) & 0.689 (0.021) & 0.782 (0.011) \\
        BERT-base & 0.912 (0.016) & 0.824 (0.032) \\
        SciBERT-base & 0.910 (0.010) & 0.819 (0.021) \\
        \bottomrule
    \end{tabular}
    \caption{Model performance averaged over five folds, shown with standard deviations in parentheses.}
    \label{tab:classifier}
\end{table}

Baseline performance of these models suggests that they are reasonably good at identifying the correct semantic levels present in an alt text sentence. Performance could be improved further through additional model tuning or data annotation. The outputs of these models can feasibly be used to provide feedback to authors who are writing alt text, to indicate when content of certain semantic levels may be lacking. 

In addition to helping to improve author-written alt text, such a classifier could be used to support improved reading experiences of existing alt text. For example, semantic level classification could enable users to make informed decisions about whether and how to read alt text by filtering for semantic levels that may be more relevant to their needs.
\citet{Morash2015GuidingNW} described similar types of personalized reading experiences, which could be enabled for alt text that has been written using standardized templates.
 We leave the implementation of these authoring and reading interfaces, as well as explorations on user interface design, to future work.

\subsection{Training and evaluating NLP models for alt text generation}
\label{sec:generation}

Recent developments in multimodal image-language pretraining \citep{Lu2019ViLBERTPT, Zhou2020UnifiedVP, Li2019VisualBERTAS} hold promise towards the eventual automatability of scientific figure alt text generation. Currently, figure alt text generation is hampered by the lack of realistic training and evaluation data. Though the size of our dataset is small and insufficient for training neural models, it may still be useful to help scale the collection of training data. Alt text from this dataset can be used to provide high-quality examples to annotators. Additionally, a classifier trained to predict alt text semantic levels such as the one introduced in Section~\ref{sec:classifier} could be used to offer feedback to annotators during the annotation process, e.g., by indicating when information of a certain content level is missing in the annotation. Instruction specific to the missing level could be provided to the annotator to elicit further description information, as in the techniques employed by \citet{Morash2015GuidingNW}. Several works propose a hybrid approach that combines machine learning model output with human writers to create better image descriptions \citep{qian-formative, Mack2021DesigningTF, Gurari2018VizWizGC}. Our dataset and classifiers could be used in collaboration with human writers to produce more descriptive alt text.

The alt texts in our dataset could also serve as part of a viable evaluation corpus. Though not all alt text in the dataset contain semantic level annotations, the texts themselves are written by the original paper authors, and are therefore more likely to be faithful to the original intent and content of the paper. We release all \numvalidalttext valid alt text extracted from our sample of papers, which includes alt text in addition to the \totalalttext alt texts belonging to data visualizations which we annotate for semantic content. These alt texts can be used to assess pieces of information that authors found important enough to include in the image description. The output of a general-purpose scientific alt text generation model can be evaluated against the information contained in the original author-written alt text associated with these figures.

\section{Discussion \& Conclusion}
\label{sec:discussion}

In regards to scientific alt text, availability is still the primary issue. However, in the alt text we were able to extract from HCI publications, we observed that for papers where authors have taken the time to write alt text, the content and level of detail available in these alt text is also worth considering. What does it mean to write useful alt text? What does it mean to include enough detail such that the content of an image can be understood by a BLV user? For graphs and charts, we propose that authors leverage the framework introduced in \citet{Lundgard2022AccessibleVV} to ensure that some basic semantic information is provided, enough such that BLV users can understand the structure of the graph, its extrema and outliers, as well as the obvious trends and comparisons that can be made. In our current analysis, we find that many author-written alt text are not yet meeting these thresholds.

We recognize the limitations of our techniques. The alt text and figures included in our dataset make up a biased sample, containing only papers from CHI and ASSETS and from the authors publishing in these venues. They are not representative of figures in all scholarly documents. Though we would have liked to construct a more representative sample of scientific figures, the overwhelming lack of figure alt text in scholarly publications prevents us from doing so. As more authors from other disciplines begin including alt text and the barriers to adding alt text to scientific figures decreases, we hope that it will become easier to create such a dataset.

Additionally, our analysis and annotations are limited to figures containing data visualizations (graphs, charts, and plots), which are only one of many types of images present in scientific publications. Our results on the suitability or missingness of semantic content in author-written alt text cannot generalize beyond these image types. Going beyond graphs, frameworks other than that introduced in \citet{Lundgard2022AccessibleVV} may be needed to capture the availability and distribution of informational content. We emphasize that the lack of certain semantic levels in alt text is not equivalent to an assessment about the alt text's quality. Different users may want different information out of alt text, and there is no one-size-fits-all solution \citep{Mack2021DesigningTF}. Rather, we use the framework as a proxy for content availability, which can be used to elicit different kinds of descriptive information that may be missing in an author's original alt text. Finally, in \citet{Lundgard2022AccessibleVV}, the authors assessed whether the levels were useful for BLV users, but not whether all of levels 1--3 were necessary for an alt text to be considered complete. Further study is needed to determine the appropriate balance of depth of information, completeness, and brevity in relation to the usefulness of alt text.

We propose in Section~\ref{sec:uses} several uses of this dataset towards improving author and publisher workflows around writing figure alt text. A classifier trained to detect semantic levels can be used to provide feedback to authors who are writing alt text, and can be used to elicit alt text containing more levels of information. The dataset can also be used to help develop machine learning models that can generate better alt text based on the image itself. We believe that coupling machine learning models with crowdsourced image descriptions may provide a reasonable solution to problems around alt text availability, and we plan to explore such solutions in future work.

Policy clearly matters. We faced significant challenges when attempting to extract alt text from a broad swathe of scientific publications, and had to limit ourselves ultimately to HCI publishing venues such as ASSETS and CHI. There is no doubt in our minds that the alt text we were able to extract are only there because of the efforts of the accessibility and HCI research community and the importance that members have placed on digital accessibility. Significant work remains to encourage researchers outside of these communities to participate in making their work accessible. Within the community, there are also ways that we can improve figure accessibility, by providing information that are described by BLV users as being more relevant or more important towards interpreting these images.

\begin{acks}

We thank Donal Fitzpatrick, Doug Downey, the reviewers, and members of the Semantic Scholar research team for their valuable feedback. 
We thank Bailey Kuehl and Ihsan Allah Rakha for their assistance in the annotation process.

\end{acks}

\bibliographystyle{ACM-Reference-Format}
\bibliography{refs}

\appendix

\section{Heuristics for identifying graphs and charts}
\label{app:heuristic_terms}

The full list of words and phrases used to identify figures corresponding to graphs and charts include:

\begin{itemize}
    \item graph
    \item chart
    \item plot
    \item scatter plot
    \item scatter
    \item distribution
    \item data
    \item points
    \item error
    \item error bar
    \item trial
    \item trials
    \item bar plot
    \item bar
    \item venn
    \item mean
    \item average 
\end{itemize}

\end{document}